\documentclass{article}%
\usepackage{amsfonts}
\usepackage{amssymb}
\usepackage{amsmath}
\usepackage{graphicx}%
\providecommand{\U}[1]{\protect\rule{.1in}{.1in}}

\newtheorem{theorem}{Theorem}

\begin{document}

\title{New Light on the Objective Indefiniteness or Literal Interpretation of Quantum Mechanics}
\author{David Ellerman\\University of California Riverside\\University of Ljubljana, Slovenia}
\maketitle

\begin{abstract}
\noindent The development of the new logic of partitions (= equivalence
relations) dual to the usual Boolean logic of subsets, and its quantitative
version as the new logical theory of information provide the basic
mathematical concepts to describe distinctions/indistinctions,
definiteness/indefiniteness, and distinguishability/indistinguishability. They
throw some new light on the objective indefiniteness or literal interpretation
of quantum mechanics (QM) advocated by Abner Shimony. This paper shows how the
mathematics of QM is the math of indefiniteness and thus, literally and
realistically interpreted, it describes an objectively indefinite reality at
the quantum level. In particular, the mathematics of wave propagation is shown
to \textit{also} be the math of the evolution of indefinite states that do not
change the degree of indistinctness between states. This corrects the
historical wrong turn of seeing QM as \textquotedblleft wave
mechanics\textquotedblright\ rather than the mechanics of particles with
indefinite/definite properties. For example, the so-called \textquotedblleft
wave-particle duality' for particles is the juxtaposition of the evolution of
a particle having an indefinite position (\textquotedblleft wave-like"
behavior) with a particle having a definite position (particle-like behavior).

\end{abstract}
\tableofcontents

\section{Introduction}

New developments in mathematical logic and the related logical information
theory have helped to further elucidate what Abner Shimony advocated as the
objective indefiniteness or literal interpretation of quantum mechanics (QM).

\begin{quotation}
\noindent From these two basic ideas alone -- indefiniteness and the
superposition principle -- it should be clear already that quantum mechanics
conflicts sharply with common sense. If the quantum state of a system is a
complete description of the system, then a quantity that has an indefinite
value in that quantum state is objectively indefinite; its value is not merely
unknown by the scientist who seeks to describe the system. Furthermore, since
the outcome of a measurement of an objectively indefinite quantity is not
determined by the quantum state, and yet the quantum state is the complete
bearer of information about the system, the outcome is strictly a matter of
objective chance -- not just a matter of chance in the sense of
unpredictability by the scientist. Finally, the probability of each possible
outcome of the measurement is an objective probability. \cite[p.
47]{shim:reality}

These statements ... may collectively be called \textquotedblleft the Literal
Interpretation\textquotedblright\ of quantum mechanics. This is the
interpretation resulting from taking the formalism of quantum mechanics
literally, as giving a representation of physical properties themselves,
rather than of human knowledge of them, and by taking this representation to
be complete. \cite[pp. 6-7]{shim:vienna}
\end{quotation}

\noindent The same theme has been continued by Shimony's student and
colleague, Gregg Jaeger.

\begin{quotation}
\noindent The conceptual elements of quantum theory that now underlie our
picture of the physical world include objective chance, quantum interference,
and the objective indefiniteness of dynamical quantities. Quantum
interference, which is directly observable, was readily absorbed by the
physics community. Objective chance and indefiniteness, being of more
philosophical significance, gained acceptance only after much debate and
conceptual analysis, when it was recognized that observed phenomena are better
understood through these notions than through older ones or hidden variables.
\cite[p. vii]{jaeger:2014}
\end{quotation}

\noindent Since the elucidation of this interpretation of QM pivots on the
notions of indefiniteness and indistinguishability, it will be called the
\textit{objective indefiniteness (OI) interpretation}.

\section{Partition logic and logical information theory}

In the past, the most basic form of logic was the Boolean logic of subsets
(usually called \textquotedblleft propositional\textquotedblright\ logic), but
from the mathematical point of view, it is only half of logic. The notion of a
subset has a category-theoretic dual in the notion of a quotient set,
equivalence relation, or partition (three equivalent notions). With some
anticipation in the work of Gian-Carlo Rota \cite{finberg:rota}, the logic of
partitions was developed for arbitrary partitions on a set
(\cite{ell:partitionlogic}; \cite{ell:intropartlogic})--so that, at the
mathematical level, partition logic is equally fundamental as the logic of the
dual notion, the usual Boolean logic of subsets.

A \textit{partition} $\pi=\left\{  B_{1},...,B_{m}\right\}  $ on a set
$U=\left\{  u_{1},...,u_{n}\right\}  $ is a set of non-empty subsets or
\textit{blocks} $B_{j}\subseteq U$ that are disjoint and whose union is
$U$.\footnote{Partition logic works with arbitrary sets but, for expository
purposes, we restrict ourselves here to finite sets and finite dimensional
Hilbert spaces.} A (real-valued) \textit{numerical attribute} on $U$ is a
function $f:U\rightarrow%
\mathbb{R}
$. It has certain numerical values, say $\left\{  r_{1},...,r_{m}\right\}  $,
and its inverse-image is a partition $\pi=\left\{  f^{-1}\left(  r_{j}\right)
\right\}  _{j=1,...,m}$ on $U$ with blocks $B_{j}=f^{-1}\left(  r_{j}\right)
$. Partitions are important for the objective indefiniteness interpretation of
QM because they explicate the notions of indistinction (or indefiniteness or
indistinguishability) and distinction at the logical level. Two elements
$u,u^{\prime}\in B_{j}=f^{-1}\left(  r_{j}\right)  $ are indistinct in terms
of the attribute $f$ or, equivalently, the partition $\pi$, and two elements
of $U$ in different blocks are distinct in terms of $f$ or $\pi$.

In view of the parallelism between subset logic and partition logic, each has
a quantitative version for finite $U$. The normalized number of elements
$\frac{\left\vert S\right\vert }{\left\vert U\right\vert }$ of a subset
$S\subseteq U$ is the (Laplace-Boole) \textit{probability of the event }$S$
\cite{boole:lot}. A ordered pair $\left(  u,u^{\prime}\right)  $ of elements
that are indistinct in $\pi$ is an \textit{indistinction} or \textit{indit of
}$\pi$, and the set of indits of a partition is its \textit{indit-set}
$\operatorname{indit}\left(  \pi\right)  $--which is just the equivalence
relation $\operatorname{indit}\left(  \pi\right)  \subseteq U\times U$
associated with the partition $\pi$. Similarly an ordered pair $\left(
u,u^{\prime}\right)  $ of elements in different blocks of $\pi$ is a
\textit{distinction} or \textit{dit of }$\pi$, and the set of dits of a
partition is its \textit{dit-set} $\operatorname{dit}\left(  \pi\right)
=U\times U-\operatorname{indit}\left(  \pi\right)  $ which is called an
\textit{apartness relation} or \textit{partition relation}.

The partition logic analogue of the normalized number of elements
$\frac{\left\vert S\right\vert }{\left\vert U\right\vert }$ is the normalized
number of dits $\frac{\left\vert \operatorname{dit}\left(  \pi\right)
\right\vert }{\left\vert U\times U\right\vert }$ of a partition $\pi$ which is
the \textit{logical entropy of the partition }$\pi$ \cite{ell:countingdits}.
Thus the dual to the logical (i.e., finite discrete) probability theory that
arises as the quantitative version of subset logic is the \textit{logical
theory of information} that arises as the quantitative version of partition
logic.\footnote{Both logical probability and logical entropy have obvious
generalizations when the points of $U$ have probabilties assigned to them
instead of being equiprobable.} Logical information theory is the foundational
theory of information based on the intuitive idea of information as
distinctions, differences, and distinguishability \cite{ell:newfound}. All the
usual definitions of simple, joint, conditional, and mutual Shannon entropy
are obtained by a uniform (dit to bit) transformation of the corresponding
definitions for logical entropy. The intuitive idea is that instead of
counting the number of distinctions, the Shannon entropy counts the (average
minimum) number of letters in a binary code (bits) it takes to make the same
distinctions (i.e., to uniquely encode the distinct messages), so the Shannon
theory is repositioned as the specialized theory about coding and
communications \cite{shannon:belljournal}.

Our purpose is to show how these new developments in mathematical logic and
information theory elucidate the objective indefiniteness interpretation of
QM. Classically, reality was thought to be \textquotedblleft definite all the
way down.\textquotedblright\ But QM\ gives a different message--that reality
is (objectively) indefinite at the quantum level. The problem is that we have
little idea how to intuitively imagine such a reality and hence the problem of
building a realistic interpretation of QM that goes beyond the bare
mathematical formalism. Since the development of QM in the first quarter of
the twentieth century, interpretations have multiplied rather than converged
which indicates the difficulty of the problem. But if quantum reality is
objectively indefinite, then this new mathematics built on the notions of
distinction/indistinction, definiteness/indefiniteness, and
distinguishability/indistinguishability, will provide some important tools to
elucidate that reality.

\section{What is a superposition state?}

\subsection{The two classical notions of abstraction}

The strategy of elucidation is to consider certain classical concepts, such as
a set of elements $S\subseteq U$ which could be a subset $S\subseteq B$ of a
block in a partition, and then to show how its features are vastly generalized
in the the corresponding quantum concept, such as a superposition state.

To understand the classical version of a superposition state, we need to
consider the abstraction principle which is most clearly understood in
mathematics. If $\sim$ is an equivalence relation on $U$ and $A\left(
u\right)  $ is \textquotedblleft the abstraction from $u$\textquotedblright,
then the\textit{ abstraction principle} turns equivalence into identity:

\begin{center}
$A\left(  u\right)  =A\left(  u^{\prime}\right)  $ iff $u\sim u^{\prime}$
\end{center}

\noindent for all $u,u^{\prime}\in U$. A well-known example of an abstraction
principle is Frege's \textquotedblleft direction principle\textquotedblright%
\ which Stewart Shapiro described as: for any lines $l_{1}$ and $l_{2}$ in
some domain, the \textquotedblleft direction of $l_{1}$ is identical to the
direction of $l_{2}$ if and only if $l_{1}$ is parallel to $l_{2}%
$.\textquotedblright\ \cite[p. 107]{shapiro:cut} Abstraction turns equivalence
of being parallel into the identity of direction. But there are two different
ways for this abstraction principle to be satisfied. The version often used by
the proverbial `working mathematician' will be called \textit{the }%
$\#1$\textit{ abstraction}, namely, just the equivalence class, e.g., a block
in the partition of a set of lines in a plane where the equivalence is
\textquotedblleft$l_{1}$ is parallel to $l_{2}$\textquotedblright. If $\left[
l\right]  $ is the parallelism equivalence class of the line $l$, then the
abstraction principle of turning equivalence into identity is clearly
satisfied: $l_{1}\simeq l_{2}$ iff $\left[  l_{1}\right]  =\left[
l_{2}\right]  $ (where $\simeq$ is the equivalence relation of being parallel).

But there is a second way to interpret abstraction and that is the one
relevant to understanding superposition in QM. It will be referred to as
\textit{the }$\#2$\textit{ type of abstraction} where the \textquotedblleft
the direction of $l$\textquotedblright\ is an abstract object that is definite
on what is common to parallel lines (i.e., their direction) but
\textit{abstracts away from where they differ, i.e., is indefinite on how they
differ}.

Within mathematics, the $\#2$ type of abstraction is highlighted by the recent
development of homotopy type theory. There is an equivalence relation $A\simeq
B$ between topological spaces which is realized by a continuous map
$f:A\rightarrow B$ such that there is an inverse $g:B\rightarrow A$ so the
$fg:B\rightarrow B$ is homotopic to $1_{B}$ (i.e., can be continuously
deformed in $1_{B}$) and $gf$ is homotopic to $1_{A}$. According to the
`classical' homotopy theorist, Hans-Joachim Baues, \textquotedblleft Homotopy
types are the equivalence classes of spaces\textquotedblright\ \cite{baues:ht}
under this equivalence relation. That is the $\#1$ type of abstraction.

But the interpretation offered in homotopy type theory (HoTT) is expanding
identity to \textquotedblleft coincide with the (unchanged) notion of
equivalence\textquotedblright\ in the words of the Univalent Foundations
Program \cite[p. 5]{ufp:hott} so it would refer to the $\#2$ homotopy type,
i.e., `\textit{the} homotopy type' that is definite on the mathematical
properties shared by all spaces in an equivalence class of homotopic spaces
(but is indefinite on the differences). Expanding identity to coincide with
equivalence is another way to describe the $\#2$ abstracting from the class
$S$ of equivalent entities to the abstract entity that is definite on what is
common to the elements $u\in S$ but is indefinite on where they differ.

For instance, `\textit{the} homotopy type' is not one of the classical
topological spaces (with points etc.) in the $\#1$ equivalence class of
homotopic spaces--just as Frege's $\#2$ abstraction of direction is not among
the lines in the equivalence class of parallel lines with the \textit{same} direction.

\begin{quotation}
\noindent\noindent While classical homotopy theory is analytic (spaces and
paths are made of points), homotopy type theory is synthetic: points, paths,
and paths between paths are basic, indivisible, primitive notions. \cite[p.
59]{ufp:hott}
\end{quotation}

Consider the homotopy example of `\textit{the} path going once (clockwise)
around the hole' in an annulus $A$ (disk with one hole as in Figure 1), i.e.,
the abstract entity $1$ in the fundamental group $\pi_{0}\left(  A\right)  $
of the annulus: $1\in\pi_{0}\left(  A\right)  \cong%
\mathbb{Z}
$:%

\begin{center}
\includegraphics[
height=1.7469in,
width=1.407in
]%
{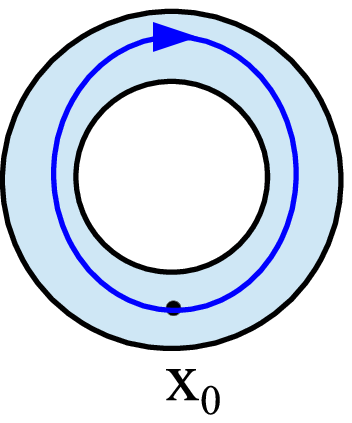}%
\end{center}

\begin{center}
Figure 1: `\textit{the} path going once (clockwise) around the hole'
\end{center}

\noindent Note that `\textit{the} path going once (clockwise) around the hole'
has the definite property of \textquotedblleft going once (clockwise) around
the hole\textquotedblright\ but is indefinite on any of the particular
(coordinatized) paths that constitute the equivalence class of coordinatized
once-around paths deformable into one another.

In a similar manner, we can view other common $\#2$ abstractions such as:
`\textit{the} cardinal number $5$' that captures what is common to the
isomorphism class of all five-element sets; `\textit{the} integer $1$
$\operatorname{mod}\left(  n\right)  $' that captures what is common within
the equivalence class $\left\{  ...,-2n+1,-n+1,1,n+1,2n+1,...\right\}  $ of
integers; `\textit{the} circle' or `\textit{the} equilateral triangle'--and so
forth.\footnote{Category theory helped to motivate homotopy type theory for
good reason. Category theory has no notion of identity between objects, only
isomorphism as `equivalence' between objects. Therefore category theory can be
seen as a theory of \textit{abstract} $\#2$ objects (i.e., the $\#2$ abstract
of an isomorphism class), e.g., abstract sets, groups, spaces, etc.}

The notion of an entity that is partly indefinite and partly definite might be
intuitively clarified by considering the difference between the two ways that
police get a picture of a suspect, a mugbook and a police artist using a
sketchpad. The mugbook is a set of definite images akin to the classical
physics notion of reality.\ But a police sketch artist starts with an
indefinite face on the sketchpad and then builds up more definiteness--which
is akin to the notion of an indefinite quantum reality that is made more
definite by a series of (compatible) measurements.

\subsection{The general notion of \#2 abstraction}

The type $2$ abstraction is usually applied in mathematics to the elements of
an equivalence class but we can apply the idea to any arbitrary (nonempty)
subset $S\subseteq U$ to arrive at the idea of an abstract entity $u_{S}$ that
is definite on what is common to the elements of $S$ and indefinite on where
they differ. In terms of the abstraction principle: $u_{S}\left(  u\right)
=u_{S}\left(  u^{\prime}\right)  $ iff $u,u^{\prime}\in S$, for all
$u,u^{\prime}\in U$. Intuitively, we might use the crutch of thinking of
$u_{S}$ as resulting from blobbing, blurring, or smearing together the
elements of $S$ to obtain $u_{S}$--so the only definite characteristics left
in $u_{S}$ are the common properties and the properties where the elements of
$S$ differ are blurred out as indefinite. But we need a more exact way to
specify the difference between $S$ and $u_{S}$.

The notion of the \textit{incidence matrix} $I\left(  R\right)  $ of a binary
relation $R\subseteq U\times U$ on $U$ supplies the right mathematical notion
to distinguish $S$ and $u_{S}$; it is the $n\times n$ matrix with rows and
columns corresponding to the elements $u_{1},...,u_{n}\in U$ such that:

\begin{center}
$I\left(  R\right)  _{ij}=\left\{
\begin{array}
[c]{l}%
1\text{ if }\left(  u_{i},u_{j}\right)  \in R\\
0\text{ otherwise.}%
\end{array}
\right.  $
\end{center}

\noindent Then the set $S\subseteq U$ of distinct elements $u_{i}\in S$ could
be represented by the incidence matrix $I\left(  \Delta S\right)  $ of the
binary relation $\Delta S=\left\{  \left(  u_{i},u_{i}\right)  :u_{i}\in
S\right\}  $ whose only non-zero elements are the diagonal elements of $1$
corresponding to the $u_{i}\in S$. Then the \textquotedblleft
blobbed-out\textquotedblright\ or \textquotedblleft blurred\textquotedblright%
\ version $u_{S}$ abstracted from $S$ would be represented by the incidence
matrix $I\left(  S\times S\right)  $ with the entries $I\left(  S\times
S\right)  _{ij}=1$ if $u_{i},u_{j}\in S$ and $0$ otherwise. The non-zero
off-diagonal elements $I\left(  S\times S\right)  _{ij}=1$ for $i\neq j$
indicated that $u_{i},u_{j}\in S$ are blobbed or blurred together or `cohere'
together in the entity $u_{S}$.

\subsection{From incidence to density matrices}

If the incidence matrices $I\left(  \Delta S\right)  $ and $I\left(  S\times
S\right)  $ are normalized by dividing through by their trace $\left\vert
S\right\vert $, then we obtain two \textit{density matrices} denoted:

\begin{center}
$\rho\left(  \Delta S\right)  =\frac{1}{\operatorname{tr}\left[  I\left(
\Delta S\right)  \right]  }I\left(  \Delta S\right)  $ and $\rho\left(
S\right)  =\frac{1}{\operatorname{tr}\left[  I\left(  S\times S\right)
\right]  }I\left(  S\times S\right)  $.
\end{center}

\noindent If we assign the equal probabilities to the elements of $S$:
$\Pr\left(  u_{i}\right)  =p_{i}=\frac{1}{\left\vert S\right\vert }$ if
$u_{i}\in S$ and $0$ otherwise, then the diagonal elements of $\rho\left(
\Delta S\right)  $ and $\rho\left(  S\right)  $ are the probabilities of
drawing the corresponding element from $U$.

This incidence matrix approach to density matrices can be generalized by
starting with any set of point probabilities $\Pr\left(  u_{i}\right)  =p_{i}$
for $u_{i}\in U$. Then the subset $S$ could be represented as a normalized
column vector $\left\vert S\right\rangle $ which $i^{th}$ entry is
$\sqrt{\frac{p_{i}}{\Pr\left(  S\right)  }}$ if $u_{i}\in S$ and $0$ otherwise
where $\Pr\left(  S\right)  =\sum_{u_{i}\in S}p_{i}$. Then the density matrix
$\rho\left(  S\right)  $ would be constructed as the (outer) product of the
column vector $\left\vert S\right\rangle $ times its transpose denoted
$\left\langle S\right\vert =\left\vert S\right\rangle ^{t}$:

\begin{center}
$\rho\left(  S\right)  _{ij}=\left(  \left\vert S\right\rangle \left\langle
S\right\vert \right)  _{ij}=\frac{1}{\Pr\left(  S\right)  }\sqrt{p_{i}p_{j}}$
if $u_{i},u_{j}\in S$ and $0$ otherwise.
\end{center}

\noindent Then a density matrix $\rho\left(  \pi\right)  $ can be associated
with a partition $\pi=\left\{  B_{1},...,B_{m}\right\}  $ on $U$ with the
point probabilities $p=\left\{  p_{1},...,p_{n}\right\}  $ by taking the
probability weighted sum of the density matrices for the blocks $B_{j}$ of
$\pi$:

\begin{center}
$\rho\left(  \pi\right)  =\sum_{j=1}^{m}\Pr\left(  B_{j}\right)  \rho\left(
B_{j}\right)  $.
\end{center}

\noindent Then a non-zero off-diagonal entry $\rho\left(  \pi\right)
_{ii^{\prime}}=\sqrt{p_{i}p_{i^{\prime}}}$ means that $u_{i}$ and
$u_{i^{\prime}}$ cohere together in some block $B_{j}$ and that $\left(
u_{i},u_{i^{\prime}}\right)  \in\operatorname{indit}\left(  \pi\right)  $ is
an indistinction of the partition $\pi$. Those non-zero off-diagonal entries
$\rho\left(  \pi\right)  _{ii^{\prime}}=\sqrt{p_{i}p_{i^{\prime}}}$ can be
thought of as an \textit{\textquotedblleft amplitude\textquotedblright\ for
}$u_{i}$\textit{ and }$u_{i^{\prime}}$\textit{ to cohere together} since the
square $p_{i}p_{i^{\prime}}$ is the probability that the ordered pair indit
$\left(  u_{i},u_{i^{\prime}}\right)  $ will be drawn (in that order) in two
independent draws from the sample space $U$.

There is zero coherence amplitude in $\rho\left(  \pi\right)  $ for elements
$u_{i}$ and $u_{i^{\prime}}$ in different blocks of $\pi$, i.e., for $\left(
u_{i},u_{i^{\prime}}\right)  \in\operatorname{dit}\left(  \pi\right)  $. The
most decoherent partition (with no coherence amplitudes) is the
\textit{discrete partition} $\mathbf{1}_{U}=\left\{  \left\{  u_{i}\right\}
\right\}  _{i=1,...,n}$ with all the blocks are singletons so no elements of
$U$ are blobbed or blurred together. Then $\rho\left(  \mathbf{1}_{U}\right)
$ is diagonal matrix with diagonal entries $\rho\left(  \mathbf{1}_{U}\right)
_{ii}=p_{i}$.

\subsection{Density matrices in quantum mechanics}

The transition to QM\ is rather clear. The elements $u_{i}\in U$ generalize to
the vectors $\left\vert u_{i}\right\rangle $ in an orthonormal (ON) basis
$\mathcal{U}=\left\{  \left\vert u_{i}\right\rangle \right\}  _{i=1,...,n}$
for an $n$-dimensional Hilbert space $V$. A superposition state $\left\vert
\psi\right\rangle \in V$ can be represented as a superposition of vectors in
the ON basis:

\begin{center}
$\left\vert \psi\right\rangle =\sum_{i=1}^{n}\left\langle u_{i}|\psi
\right\rangle \left\vert u_{i}\right\rangle =\sum_{i}\alpha_{i}\left\vert
u_{i}\right\rangle $
\end{center}

\noindent where $\alpha_{i}=\left\langle u_{i}|\psi\right\rangle $. The first
classical approximation to a superposition state was the blobbed-out or
blurred version $u_{S}$ of a subset $S$ which was definite on the attributes
common to the elements of $S$ and indefinite concerning the properties that
differ between the elements of $S$. This blurred version $u_{S}$ of $S$ could
be represented by the incidence matrix $I\left(  S\times S\right)  $ where two
elements $u_{i}$ and $u_{j}$ cohered or were blurred together iff they were
both in $S$. The normalized incidence matrix $\frac{1}{\operatorname{tr}%
\left[  I\left(  S\times S\right)  \right]  }I\left(  S\times S\right)  $ was
a density matrix that could be further refined by introducing different point
probabilities $p=\left\{  p_{1},...,p_{n}\right\}  $. Then we have the density
matrix $\rho\left(  S\right)  $ whose non-zero off-diagonal entries $\frac
{1}{\Pr\left(  S\right)  }\sqrt{p_{i}p_{i^{\prime}}}$ give the amplitude for
$u_{i}$ to cohere or blur together with $u_{i^{\prime}}$in $S$. $\rho\left(
S\right)  $ is our final classical representation of the $\#2$ abstraction
from $S$: from $u_{S}$ to $I\left(  S\times S\right)  $ to $\frac
{1}{\operatorname{tr}\left[  I\left(  S\times S\right)  \right]  }I\left(
S\times S\right)  $ to $\rho\left(  S\right)  $.

The quantum version of $\rho\left(  S\right)  $ is $\rho\left(  \psi\right)
=\left\vert \psi\right\rangle \left\langle \psi\right\vert $ with the entries
$\rho\left(  \psi\right)  _{ij}=\alpha_{i}\alpha_{j}^{\ast}$ (where
$\alpha_{j}^{\ast}$ is the complex conjugate of $\alpha_{j}=\left\langle
u_{j}|\psi\right\rangle $). Then the non-zero off-diagonal elements
$\alpha_{i}\alpha_{j}^{\ast}$ for $i\neq j$ give the \textit{amplitude for
}$\left\vert u_{j}\right\rangle $\textit{ to cohere or blur together with
}$\left\vert u_{j}\right\rangle $\textit{ in the superposition }$\left\vert
\psi\right\rangle $--the \textit{indistinction amplitude}, and those elements
are usually called \textquotedblleft coherences\textquotedblright\ \cite[p.
302]{cohen-tann:qm}. This blurring together of elements in the classical $\#2$
abstraction is a key characteristic in the quantum case.

\begin{quotation}
\noindent\lbrack The] off-diagonal terms of a density matrix...are often
called \textit{quantum coherences} because they are responsible for the
interference effects typical of quantum mechanics that are absent in classical
dynamics. \cite[p. 177]{auletta:qm}
\end{quotation}

It might be useful to connect this notion of a superposition state as a partly
definite and indefinite entity. i.e., that is definite only on the properties
common to the superposed states and indefinite otherwise, to common examples
such as the double-slit experiment or the Mach-Zehnder interferometer. In the
double-slit experiment, consider the superposition state $\left\vert
\psi\right\rangle =\frac{1}{\sqrt{2}}\left(  \left\vert slit1\right\rangle
+\left\vert slit2\right\rangle \right)  $. This is commonly described as the
state of the particle as \textquotedblleft going through two slits at the same
time\textquotedblright\ \cite[p. 94]{anil:twodoors}. But that assumes that
there is a definite particle that is going through each slit. But the
objectively indefinite interpretation of QM would interpret the superposition
$\left\vert \psi\right\rangle $ as the blurred-together state of being
indefinite as to which slot the particle goes through--and only being definite
on going through the slits. Abner Shimony found one of Yogi Berra's
malapropisms to be quite appropriate: \textquotedblleft If you come to a fork
in the road, take it.\textquotedblright\ \cite[p. 5]{shim:vienna} We do not
have a `clear and distinct idea' how to imagine such an indefinite
state--although many find the crutch of a definite wave hitting both slits as
being helpful (but misleading) imagery.

In the case of the Mach-Zehnder interferometer, the superposition state
$\left\vert \phi\right\rangle =\frac{1}{\sqrt{2}}\left(  \left\vert
arm1\right\rangle +\left\vert arm2\right\rangle \right)  $ (after the first
beam-splitter) is often described as the photon going through both arms. But
on the OI interpretation, it would be more accurate to say that the photon is
in the state of being indefinite (i.e., blurred) between the two arms but is
definitely going through the arms of the apparatus.\footnote{For more on these
and other apparatuses in the context of delayed-choice experiments, see
\cite{ell:delayed}}

The point might be illustrated using our mugbook-sketchpad analogy. Suppose a
witness has found two pictures of different people in the mugbook that she
thinks equally depict the suspect. That is analogous to the description of the
particle as definitely going through both slits or both arms of apparatus. But
the OI interpretation would take the proper analogy as being a partial sketch
of the suspect that is partly definite (e.g., on the characteristics common to
the two mugshots) and partly indefinite (e.g., on where the two mugshots
differ). To extend the analogy to the mathematics, such a partial sketch could
be represented as the superposition: $\left\vert \varphi\right\rangle
=\frac{1}{\sqrt{2}}\left(  \left\vert mugshot1\right\rangle +\left\vert
mugshot2\right\rangle \right)  $.

\section{Interpreting the inner product}

\subsection{The classical case}

Classically, we might take the \textit{norm of a subset} $S\subseteq U$ as
$\left\Vert S\right\Vert =\sqrt{\left\vert S\right\vert }$, the square root of
its cardinality. The amplitude of the overlap between sets $S,T\subseteq U$ is
$\left\Vert S\cap T\right\Vert =\sqrt{\left\vert S\cap T\right\vert }$ so the
square of that overlap amplitude is the cardinality $\left\vert S\cap
T\right\vert $. In terms of our leifmotif of distinction and indistinction,
$\left\Vert S\cap T\right\Vert $ measures the \textit{amplitude of
indistinction} between $S$ and $T$. The maximum value is when they are fully
indistinct, $\left\Vert S\cap T\right\Vert =\left\Vert S\cap S\right\Vert
=\left\Vert T\cap T\right\Vert $, and the minimum indistinction amplitude
$\left\Vert S\cap T\right\Vert =0$ means they have no overlap and have no
indistinctness, i.e., are fully distinct. If we compare a random drawing from
$S$ to a random drawing from $T$, then we could always distinguish between the
drawings no matter what the outcome iff $\left\Vert S\cap T\right\Vert =0$.

Given the set $U=\left\{  u_{1},...,u_{n}\right\}  $, the coefficients
$\left\Vert \left\{  u_{i}\right\}  \cap S\right\Vert $ represent the
amplitude of $\left\{  u_{i}\right\}  $'s indistinctness with $S$, and its
square $\left\Vert \left\{  u_{i}\right\}  \cap S\right\Vert ^{2}=\left\vert
\left\{  u_{i}\right\}  \cap S\right\vert $ represents the proportion of $S$
that is $\left\{  u_{i}\right\}  $. The non-zero proportions $\left\Vert
\left\{  u_{i}\right\}  \cap S\right\Vert ^{2}=\left\vert \left\{
u_{i}\right\}  \cap S\right\vert $ add up to equal $\left\Vert S\right\Vert
^{2}=\left\vert S\right\vert $. Taking the uniform probability distribution on
$U$, the probability $\Pr\left(  u_{i}|S\right)  $ of drawing a particular
element $u_{i}\in S$ is the normalized proportion $\frac{\left\Vert \left\{
u_{i}\right\}  \cap S\right\Vert ^{2}}{\left\Vert S\right\Vert ^{2}}%
=\frac{\left\vert \left\{  u_{i}\right\}  \cap S\right\vert }{\left\vert
S\right\vert }=\frac{1}{\left\vert S\right\vert }$.

\subsection{The quantum case}

The quantum version of the overlap amplitude between two states $\left\vert
\psi\right\rangle $ and $\left\vert \phi\right\rangle $ in $V$ is their inner
product $\left\langle \phi|\psi\right\rangle $ which can be interpreted as the
\textit{amplitude of their indistinctness}. They are maximally indistinct when
$\left\langle \phi|\psi\right\rangle =\left\langle \phi|\phi\right\rangle
=\left\langle \psi|\psi\right\rangle $ and have no indistinctness, i.e., are
fully distinct, when $\left\langle \phi|\psi\right\rangle =0$. When comparing
the measurement of $\left\vert \psi\right\rangle $ to the measurement of
$\left\vert \phi\right\rangle $ (using some ON measurement basis
$\mathcal{U}=\left\{  \left\vert u_{i}\right\rangle \right\}  _{i=1,...,n}$),
they are fully distinguishable regardless of the outcome iff $\left\langle
\phi|\psi\right\rangle =0$.

Given an ON basis $\mathcal{U}=\left\{  \left\vert u_{i}\right\rangle
\right\}  _{i=1,...,n}$, the coefficients $\left\langle u_{i}|\psi
\right\rangle =\alpha_{i}$ represent the amplitude of $\left\vert
u_{i}\right\rangle $'s indistinctness with $\left\vert \psi\right\rangle $,
and its absolute square $\alpha_{i}\alpha_{i}^{\ast}=\left\Vert \left\langle
u_{i}|\psi\right\rangle \right\Vert ^{2}$ represents the proportion of
$\left\vert \psi\right\rangle $ that is $\left\vert u_{i}\right\rangle
$.\footnote{We use the notation $\left\Vert \alpha\right\Vert =\sqrt
{\alpha\alpha^{\ast}}$ for the norm of a complex number $\alpha$ to avoid
notational conflict with the cardinality $\left\vert S\right\vert $ of a
subset $S$.} The non-zero proportions $\left\Vert \left\langle u_{i}%
|\psi\right\rangle \right\Vert ^{2}$ add up to equal $\left\Vert \left\langle
\psi|\psi\right\rangle \right\Vert ^{2}$. Taking the uniform probability
distribution on the interval $\left[  0,\left\Vert \left\langle \psi
|\psi\right\rangle \right\Vert ^{2}\right]  $, the probability $p_{i}$ of a
point falling in a segment of length $\left\Vert \left\langle u_{i}%
|\psi\right\rangle \right\Vert ^{2}$ is just that normalized length of the
segment $p_{i}=\frac{\left\Vert \left\langle u_{i}|\psi\right\rangle
\right\Vert ^{2}}{\left\Vert \left\langle \psi|\psi\right\rangle \right\Vert
^{2}}$, which is also the probability of getting the outcome $\left\vert
u_{i}\right\rangle $ when measuring $\left\vert \psi\right\rangle $ using
$\left\{  \left\vert u_{i}\right\rangle \right\}  _{i=1,...,n}$ as the
measurement basis, i.e., the Born rule.

\section{Numerical attributes and measurement}

\subsection{The classical case}

Given a universe set $U=\left\{  u_{1},...,u_{n}\right\}  $ with point
probabilities $p_{1},...,p_{n}$, a real-value numerical attribute on $U$ is a
function $f:U\rightarrow%
\mathbb{R}
$. The numerical values $\left\{  r_{1},...,r_{m}\right\}  $ in the image of
$f$ define a partition $f^{-1}=\left\{  B_{1},...,B_{m}\right\}  $ on $U$ by
taking $B_{j}=f^{-1}\left(  r_{j}\right)  $ for $j=1,...,m$.

The blobbed, blurred, or smeared $\#2$ abstraction version of a nonempty
subset $S\subseteq U$ is represented by the density matrix $\rho\left(
S\right)  $, which might be called a \textit{pure} density matrix since
$\rho\left(  S\right)  ^{2}=\rho\left(  S\right)  $ and thus
$\operatorname{tr}\left[  \rho\left(  S\right)  ^{2}\right]
=\operatorname{tr}\left[  \rho\left(  S\right)  \right]  =1$. Intuitively,
this `superposition' version of $S$ is definite only on the properties common
to all the elements of $S$ and is otherwise indefinite. But the
blurred-together elements of superposition $S$ might be distinguished by
classifying them according to some numerical attribute $f$. Since the
superposition version of $S$ is represented by the density matrix $\rho\left(
S\right)  $, this classification operation might represented by an operation
on the density matrix $\rho\left(  S\right)  $ to obtain an $f$-classified
density matrix $\hat{\rho}\left(  S\right)  $. The non-zero off-diagonal
elements $\rho\left(  S\right)  _{ii^{\prime}}=\frac{1}{\Pr\left(  S\right)
}\sqrt{p_{i}p_{i^{\prime}}}$ in $\rho\left(  S\right)  $ give the amplitude
for $u_{i}$ to be indistinct with $u_{i^{\prime}}$ in the superposition
version of $S$. The transformation $\rho\left(  S\right)  \rightsquigarrow
\hat{\rho}\left(  S\right)  $ is quite simple; if $f$ distinguishes $u_{i}$
and $u_{i^{\prime}}$, i.e., if $\left(  u_{i},u_{i^{\prime}}\right)  $ is a
distinction of the partition $f^{-1}$, then and only then is the indistinction
amplitude set to $0$. If $u_{i}$ and $u_{i^{\prime}}$ are not distinguished by
$f$, i.e., $u_{i}$ and $u_{i^{\prime}}$ are not only both in $S$ but are both
in some block of $f^{-1}=\left\{  f^{-1}\left(  r_{1}\right)  ,...,f^{-1}%
\left(  r_{m}\right)  \right\}  $, then the indistinction amplitude $\frac
{1}{\Pr\left(  S\right)  }\sqrt{p_{i}p_{i^{\prime}}}$ remains the same as
before. And since no element $u_{i}$ can ever be distinguished from itself by
any numerical attribute, the diagonal elements remain the same. These changes
determine the $f$-classified density matrix $\hat{\rho}\left(  S\right)  $.
Intuitively, the blurred or superposition version of $S$ represented by
$\rho\left(  S\right)  $ has a definite attribute value only if that value is
common to all the $u_{i}\in S$, i.e., for some $j$, $S\subseteq f^{-1}\left(
r_{j}\right)  =B_{j}$, and then $\rho\left(  S\right)  =\hat{\rho}\left(
S\right)  $. Otherwise the elements of the set $S$ do not share any $f$-value
so $\hat{\rho}\left(  S\right)  ^{2}\neq\hat{\rho}\left(  S\right)  $,
$\operatorname{tr}\left[  \hat{\rho}\left(  S\right)  ^{2}\right]  <1$, and
$\hat{\rho}\left(  S\right)  $ might be called a \textit{mixed} density matrix.

The transformation of matrices $\rho\left(  S\right)  \rightsquigarrow
\hat{\rho}\left(  S\right)  $ can also be specified entirely using matrix
operations. Let $P_{B_{j}}$ be the diagonal $n\times n$ matrix with diagonal
element $\left(  P_{B_{j}}\right)  _{ii}$ equal to $1$ if $u_{i}\in B_{j}$ and
$0$ otherwise so it is a projection matrix $P_{B_{j}}^{2}=P_{B_{j}}$. Then
pre- and post-multiplying $\rho\left(  S\right)  $ by the projections
$P_{B_{j}}$ for $B_{j}\in f^{-1}$ and summing has the effect of zeroing out
all the indistinction amplitudes $\frac{1}{\Pr\left(  S\right)  }\sqrt
{p_{i}p_{i^{\prime}}}$ where $u_{i}$ and $u_{i^{\prime}}$ are distinguished by
$f$. In anticipation of the quantum case, this operation on $\rho\left(
S\right)  $ to obtain the $f$-classified $\hat{\rho}\left(  S\right)  $ will
be called the \textit{classical L\"{u}ders mixture operation}:

\begin{center}
$\hat{\rho}\left(  S\right)  =\sum_{j=1}^{m}P_{B_{j}}\rho\left(  S\right)
P_{B_{j}}$.
\end{center}

If $S=U$, the universe set, then $\rho\left(  S\right)  =\rho\left(
\mathbf{0}_{U}\right)  $, the density matrix representation of the indiscrete
partition $\mathbf{0}_{U}=\left\{  U\right\}  $. The \textit{join} of two
partitions $\pi=\left\{  B_{j}\right\}  _{j=1}^{m}$ and $\sigma=\left\{
C_{k}\right\}  _{k=1}^{m^{\prime}}$, is the partition $\pi\vee\sigma$ whose
blocks are all the non-empty intersections $B_{j}\cap C_{k}\neq\emptyset$. The
join operation combines all the distinctions made by the two partitions, i.e.,
$\operatorname{dit}\left(  \pi\vee\sigma\right)  =\operatorname{dit}\left(
\pi\right)  \cup\operatorname{dit}\left(  \sigma\right)  $. The result of
classifying $\rho\left(  \mathbf{0}_{U}\right)  $ by $\pi=f^{-1}$ is the
density matrix of the join $\mathbf{0}_{U}\vee\pi=\pi$, i.e., $\hat{\rho
}\left(  \mathbf{0}_{U}\right)  =\rho\left(  \pi\right)  $. Further
classifications by other partitions on $U$ will add to the join and thus
introduce more and more distinctions until obtaining the maximally
distinguished discrete partition $\mathbf{1}_{U}=\left\{  \left\{
u_{i}\right\}  \right\}  _{i=1}^{n}$. A set of partitions $\left\{  \pi
,\sigma,...\right\}  $ such that $\pi\vee\sigma\vee...=\mathbf{1}_{U}$ might
be called a \textit{complete set of partitions on }$U$. The density matrix
$\rho\left(  \mathbf{1}_{U}\right)  $ is the diagonal matrix $\rho\left(
\Delta U\right)  $ with the probabilities $p_{i}$'s as the diagonal entries
where all the blurring effects or indistinctions between the elements of the
superposition version of $U$ have been eliminated.

\subsection{The quantum case}

Given an orthonormal basis $\mathcal{U}=\left\{  \left\vert u_{i}\right\rangle
\right\}  _{i=1,...,n}$ for the $n$-dimensional Hilbert space $V$, a
real-valued numerical attribute is a function $f:\mathcal{U}\rightarrow%
\mathbb{R}
$ with a set of image values $\left\{  \lambda_{1},...,\lambda_{m}\right\}  $.
Extending the $f$-assignment $\left\vert u_{i}\right\rangle \longmapsto
\lambda_{i}\left\vert u_{i}\right\rangle $ linearly to the whole space $V$
defines a linear operator $F:V\rightarrow V$ with eigenvectors $\mathcal{U}%
=\left\{  \left\vert u_{i}\right\rangle \right\}  _{i=1,...,n}$ and real
eigenvalues $\lambda_{1},...,\lambda_{m}$, so $F$ is a Hermitian operator,
i.e., an observable. Conversely, each Hermitian operator $F:V\rightarrow V$
has an ON basis $\mathcal{U}=\left\{  \left\vert u_{i}\right\rangle \right\}
_{i=1,...,n}$ of eigenvectors with eigenvalues $\lambda_{1},...,\lambda_{m}$
so that assigning each eigenvector its eigenvalue gives the eigenvalue
function which is a numerical attribute $f:\mathcal{U}\rightarrow%
\mathbb{R}
$.

Given a normalized superposition state $\left\vert \psi\right\rangle $, its
resolution in terms of an ON basis $\mathcal{U}$ of eigenvectors of a
Hermitian operator $F$ gives $\left\vert \psi\right\rangle =\sum_{i=1}%
^{n}\left\langle u_{i}|\psi\right\rangle \left\vert u_{i}\right\rangle
=\sum_{i}\alpha_{i}\left\vert u_{i}\right\rangle $. The density matrix
$\rho\left(  \psi\right)  =\left\vert \psi\right\rangle \left\langle
\psi\right\vert $ represented in the $\mathcal{U}$-basis has the elements
$\rho\left(  \psi\right)  _{ii^{\prime}}=\alpha_{i}\alpha_{i^{\prime}}^{\ast}$
and is a pure state density matrix where $\rho\left(  \psi\right)  ^{2}%
=\rho\left(  \psi\right)  $ and $\operatorname{tr}\left[  \rho\left(
\psi\right)  ^{2}\right]  =1$. Let $f:\mathcal{U}\rightarrow%
\mathbb{R}
$ be the eigenvalue function assigning to each eigenvector $\left\vert
u_{i}\right\rangle $ its eigenvalue where $\lambda_{1},...,\lambda_{m}$ are
the eigenvalues of $F$. The inverse images $f^{-1}\left(  \lambda_{j}\right)
$ define a set partition $f^{-1}$ on the set $\mathcal{U}$ where each block
$B_{j}=f^{-1}\left(  \lambda_{j}\right)  $ generates the eigenspace $\left[
B_{j}\right]  \subseteq V$ associated with the eigenvalue $\lambda_{j}$ for
$j=1,...,m$.\footnote{The eigenspaces $\left[  B_{j}\right]  $ form a
direct-sum decomposition of $V$. A direct-sum decomposition of a vector space
can be considered the vector-space version of a partition on a set. Since a
set-partition (or quotient set) is category-theoretically dual to a subset, a
direct-sum decomposition of a vector space is similarly dual to a subspace.
And just as the Boolean logic of subsets has the dual logic of partitions, so
the usual notion of the quantum logic of (closed) subspaces of a Hilbert space
\cite{birkvon-n:logicqm} will have a dual form in the quantum logic of
direct-sum decompositions \cite{ell:dsd-logic}.} The observable $F$, or
equivalently the eigenvalue function $f:\mathcal{U}\rightarrow%
\mathbb{R}
$, can be used to distinguish or classify the states $\left\vert
u_{i}\right\rangle $ that are blurred together in the superposition state
$\left\vert \psi\right\rangle $ with the indistinction amplitudes or
coherences $\alpha_{i}\alpha_{i^{\prime}}^{\ast}$ between the states
$\left\vert u_{i}\right\rangle $ and $\left\vert u_{i^{\prime}}\right\rangle
$. This operation of distinguishing by classifying, usually called
\textquotedblleft(projective) measurement\textquotedblright, has the same
effect on the density matrix $\rho\left(  \psi\right)  $ (represented in the
measurement basis $\mathcal{U}$) of zeroing out (or decohering) the
indistinction amplitudes $\rho\left(  \psi\right)  _{ii^{\prime}}=\alpha
_{i}\alpha_{i^{\prime}}^{\ast}$ when and only when $\left\vert u_{i}%
\right\rangle $ and $\left\vert u_{i^{\prime}}\right\rangle $ are
distinguished by $f$, i.e., have different eigenvalues. Using the same
projection matrices $P_{B_{j}}$ where $B_{j}=f^{-1}\left(  \lambda_{j}\right)
$ as in the classical case, the post-classification or post-measurement
density matrix $\hat{\rho}\left(  \psi\right)  $ is obtained by the
\textit{quantum L\"{u}ders mixture operation} \cite[p. 279]{auletta:qm}:

\begin{center}
$\hat{\rho}\left(  \psi\right)  =\sum_{j=1}^{m}P_{B_{j}}\rho\left(
\psi\right)  P_{B_{j}}$.
\end{center}

If $G:V\rightarrow V$ is another Hermitian operator on $V$ that commutes with
$\ F$, then we can take the ON basis $\mathcal{U}$ as a basis of simultaneous
eigenvectors of both $F$ and $G$. If $g:\mathcal{U}\rightarrow%
\mathbb{R}
$ is the eigenvalue function of $G$ with eigenvalues $\mu_{1},...,\mu
_{m^{\prime}}$, then $g^{-1}$ gives a set partition on $\mathcal{U}$ and the
join $f^{-1}\vee g^{-1}$ is a more refined partition on $\mathcal{U}$ where
each block $f^{-1}\left(  \lambda_{j}\right)  \cap g^{-1}\left(  \mu
_{k}\right)  $ can be characterized by the ordered pair $\left(  \lambda
_{j},\mu_{k}\right)  $ of eigenvalues. A set of commuting operators $F,G,...$
is called a \textit{complete set of commuting operators} (CSCO) if all the
blocks in the join $f^{-1}\vee g^{-1}\vee...$ are singletons so each basis
simultaneous eigenvector in $\mathcal{U}$ can be characterized by the sequence
of eigenvalues $\left(  \lambda_{j},\mu_{k},...\right)  $. If all the
compatible measurements by the observables in a CSCO have been carried out,
then the result is the completely decohered diagonal density matrix with all
the off-diagonal coherence amplitudes eliminated.

Definiteness in QM is achieved when a state such as $\left\vert u_{i}%
\right\rangle $ has a specific eigenvalue $f\left(  \left\vert u_{i}%
\right\rangle \right)  =\lambda_{j}$.\ Intuitively, in a blobbed, blurred, or
smeared state such as a superposition $\left\vert \psi\right\rangle $, it is
definite only on the attributes that are common to all the $\left\vert
u_{i}\right\rangle $ `\textit{in}' $\left\vert \psi\right\rangle $ (in the
sense that $\left\langle u_{i}|\psi\right\rangle \neq0$), and indefinite
otherwise. In more precise terms, a superposition state $\left\vert
\psi\right\rangle $ has the definite $F$-observable value of $\lambda_{j}$ if
and only if all the $\left\vert u_{i}\right\rangle $ in the superposition
$\left\vert \psi\right\rangle $ also have that same value $\lambda_{j}$--in
which case $\hat{\rho}\left(  \psi\right)  =\rho\left(  \psi\right)  $ and
$\left\vert \psi\right\rangle \in\left[  B_{j}\right]  $, i.e., $\left\vert
\psi\right\rangle $ is one of the eigenvectors for $\lambda_{j}$. Otherwise,
the $\left\vert u_{i}\right\rangle $ in $\left\vert \psi\right\rangle $ have
no $F$-value in common so $\hat{\rho}\left(  \psi\right)  \neq\rho\left(
\psi\right)  $, $\operatorname{tr}\left[  \hat{\rho}\left(  \psi\right)
^{2}\right]  <1$, and $\hat{\rho}\left(  \psi\right)  $ is the density matrix
of a mixture.

A more non-trivial example of a partly definite and partly indefinite state is
the definite correlation obtained in an entangled superposition. Suppose an
observable $A$ can have two eigenstates $\left\vert a_{1}\right\rangle $ and
$\left\vert a_{2}\right\rangle $ in a Hilbert state $H$ and an observable $B$
has two eigenstates $\left\vert b_{1}\right\rangle $ and $\left\vert
b_{2}\right\rangle $ in another Hilbert space $H^{\prime}$. Then in the tensor
product $H\otimes H^{\prime}$, we have the definite states $\left\vert
s\right\rangle =\left\vert a_{1}\right\rangle \otimes\left\vert b_{1}%
\right\rangle $ and $\left\vert s^{\prime}\right\rangle =\left\vert
a_{2}\right\rangle \otimes\left\vert b_{2}\right\rangle $, but the entangled
superposition state $F=\frac{1}{\sqrt{2}}\left(  \left\vert a_{1}\right\rangle
\otimes\left\vert b_{1}\right\rangle +\left\vert a_{2}\right\rangle
\otimes\left\vert b_{2}\right\rangle \right)  $ is not definitely in either state.

\begin{quotation}
\noindent When the composite system is in the state $F$, however, neither $A$
nor $B$ has a definite value, but there is a definite correlation of $A$ and
$B$: $A$ and $B$ are actualized jointly either as $(a_{1},b_{1})$ or as
$(a_{2},b_{2})$. The composite system has a definite property, which can
loosely be called \textquotedblleft sameness of the indices of the possible
values of A and B,\textquotedblright\ not inferrable from the entire
specification of $s$ by itself and the entire specification of $s^{\prime}$ by
itself. \cite[p. 7]{shim:vienna}
\end{quotation}

\section{Logical information theory at the classical and quantum level}

\subsection{The classical case}

The strategy of elucidating the objective indefiniteness interpretation of QM
is to use the notions distinction and indistinction, distinguishability and
indistinguishability, first in the classical case, where they are more easily
understood, and then to recapitulate them in the quantum case. The notion of
logical entropy at the classical and quantum level captures quantitatively the
creation of distinctions from indistinctions in classification and measurement.

Given a set partition $\pi=\left\{  B_{1},...,B_{m}\right\}  $ on a set
$U=\left\{  u_{1},...,u_{n}\right\}  $, the set of distinctions or dits of
$\pi$ is the set $\operatorname{dit}\left(  \pi\right)  \subseteq U\times U$
of all ordered pairs $\left(  u_{i},u_{i^{\prime}}\right)  $ with $u_{i}$ and
$u_{i^{\prime}}$ in different blocks of $\pi$. If all the points of $U$ are
equiprobable, i.e., $p_{i}=\frac{1}{\left\vert U\right\vert }$, then the
\textit{logical entropy of }$\pi$, denoted $h\left(  \pi\right)  $, is the
normalized count of the distinctions of $\pi$, i.e., $h\left(  \pi\right)
=\frac{\left\vert \operatorname{dit}\left(  \pi\right)  \right\vert
}{\left\vert U\times U\right\vert }$. With $\Pr\left(  B_{j}\right)
=\sum\left\{  p_{i}:u_{i}\in B_{j}\right\}  $ and the complementary
equivalence relation $\operatorname{indit}\left(  \pi\right)  =U\times
U-\operatorname{dit}\left(  \pi\right)  =\cup_{j=1}^{m}B_{j}\times B_{j}$, we
can express the logical entropy as:

\begin{center}
$h\left(  \pi\right)  =\frac{\left\vert \operatorname{dit}\left(  \pi\right)
\right\vert }{\left\vert U\times U\right\vert }=\frac{\left\vert U\times
U-\cup_{j}B_{j}\times B_{j}\right\vert }{\left\vert U\times U\right\vert
}=1-\sum_{j}\frac{\left\vert B_{j}\times B_{j}\right\vert }{\left\vert U\times
U\right\vert }=1-\sum_{j}\left(  \frac{\left\vert B_{j}\right\vert
}{\left\vert U\right\vert }\right)  ^{2}=1-\sum_{j=1}^{m}\Pr\left(
B_{j}\right)  ^{2}$.
\end{center}

\noindent When $U$ has point probabilities $p_{1},...,p_{n}$, then the natural
definition is: $h\left(  \pi\right)  =1-\sum_{j=1}^{m}\Pr\left(  B_{j}\right)
^{2}$. The logical entropy of a partition $h\left(  \pi\right)  $ has the
simple interpretation: in two independent draws (i.e., with replacement) from
$U$, $h\left(  \pi\right)  $ is the probability of drawing a distinction of
$\pi$--and $\sum_{j}\Pr\left(  B_{j}\right)  ^{2}$ is the complementary
probability of drawing an indistinction of $\pi$.

The two extreme partitions are the \textit{indiscrete partition} (or `blob')
$\mathbf{0}_{U}=\left\{  U\right\}  $ which makes no distinctions so $h\left(
\mathbf{0}_{U}\right)  =0$, and the \textit{discrete partition} $\mathbf{1}%
_{U}=\left\{  \left\{  u_{i}\right\}  \right\}  _{i=1,...,n}$ which
distinguishes all the elements of $U$ so $h\left(  \mathbf{1}_{U}\right)
=1-\sum_{i=1}^{n}p_{i}^{2}$. The maximum logical entropy occurs in the
equiprobable case of $p_{i}=\frac{1}{\left\vert U\right\vert }=\frac{1}{n}$
when $h\left(  \mathbf{1}_{U}\right)  =1-\frac{1}{n}$ which is the two-draw
probability of drawing distinct elements of $U$.

The definitions are easily reformulated in terms of the density matrix
representation of the partition as: $\rho\left(  \pi\right)  =\sum_{j=1}%
^{m}\Pr\left(  B_{j}\right)  \rho\left(  B_{j}\right)  $. Then the equivalent
definition of the logical entropy $h\left(  \pi\right)  $ of $\pi$ is:

\begin{center}
$h\left(  \rho\left(  \pi\right)  \right)  =1-\operatorname{tr}\left[
\rho\left(  \pi\right)  ^{2}\right]  $.
\end{center}

\noindent A pure density matrix $\rho\left(  S\right)  $ representing the
superposition-version of $S\subseteq U$ has $\rho\left(  S\right)  ^{2}%
=\rho\left(  S\right)  $ and all density matrices have trace $1$ so the
logical entropy of pure density matrices is always zero: $h\left(  \rho\left(
S\right)  \right)  =1-\operatorname{tr}\left[  \rho\left(  S\right)
^{2}\right]  =1-1=0$. We have seen that the classification or distinguishing
of the blobbed-together elements of $S$ by a partition $\pi$ transforms the
pure density matrix $\rho\left(  S\right)  $ into the mixed density matrix
$\hat{\rho}\left(  S\right)  $ obtained by the classical L\"{u}ders mixture
operation $\hat{\rho}\left(  S\right)  =\sum_{j=1}^{m}P_{B_{j}}\rho\left(
S\right)  P_{B_{j}}$. The classification zeros all the off-diagonal
indistinction-amplitude terms $\frac{1}{\Pr\left(  S\right)  }\sqrt
{p_{i}p_{i^{\prime}}}$ for $u_{i},u_{i^{\prime}}\in S$ where $u_{i}$ and
$u_{i^{\prime}}$ are in different blocks of $\pi$. Logical entropy captures
these distinctions made by the classification of $\rho\left(  S\right)  $ by
the partition $\pi$. The fundamental theorem relating logical entropy and
classification is:

\begin{theorem}
The sum of the squares of all indistinction-amplitudes zeroed in the
L\"{u}ders mixture operation taking $\rho\left(  S\right)  $ to $\hat{\rho
}\left(  S\right)  $ is the logical entropy $h\left(  \hat{\rho}\left(
S\right)  \right)  $. \cite{ell:newfound}
\end{theorem}

\subsection{The quantum case}

The quantum case is a straight-forward generalization of the classical case.
The \textit{quantum logical entropy} of any quantum state given by a density
matrix $\rho$ is defined by:

\begin{center}
$h\left(  \rho\right)  =1-\operatorname{tr}\left[  \rho^{2}\right]  $.
\end{center}

Let $\mathcal{U}=\left\{  \left\vert u_{i}\right\rangle \right\}
_{i=1,...,n}$ again be an orthonormal basis of eigenvectors of a Hermitian
operator $F:V\rightarrow V$ with eigenvectors $\left\{  \lambda_{j}\right\}
_{j=1,...,m}$. Let $\left\vert \psi\right\rangle =\sum_{i}\alpha_{i}\left\vert
u_{i}\right\rangle $ be a normalized superposition state with $\rho\left(
\psi\right)  =\left\vert \psi\right\rangle \left\langle \psi\right\vert $ so
that $h\left(  \rho\left(  \psi\right)  \right)  =0$. The measurement of
$\left\vert \psi\right\rangle $ by the observable $F$ transforms the pure
state density operator $\rho\left(  \psi\right)  $ into the mixed state
density operator given by the L\"{u}ders mixture operation: $\hat{\rho}\left(
\psi\right)  =\sum_{j=1}^{m}P_{j}\rho\left(  \psi\right)  P_{j}$ where $P_{j}$
is the projection to the eigenspace of the eigenvalue $\lambda_{j}$. Then the
quantum logical entropy $h\left(  \hat{\rho}\left(  \psi\right)  \right)
=1-\operatorname{tr}\left[  \hat{\rho}\left(  \psi\right)  ^{2}\right]  $ has
the simple interpretation of the being the probability in two independent
$F$-measurements of $\left\vert \psi\right\rangle $ of getting different
eigenvalues. Moreover, the fundamental theorem relating quantum logical
entropy to (projective) measurement gives the detailed connection to the
changes in the density matrix $\rho\left(  \psi\right)  $ when represented in
the $U$-basis.\footnote{Both quantum logical entropy and the Von Neumann
entropy $S\left(  \rho\left(  \psi\right)  \right)  =-\rho\left(  \psi\right)
\log\left(  \rho\left(  \psi\right)  \right)  $ usually considered in QM have
the value of $0$ for pure states and increase under (projective) measurement.
But there seems to be no similar relation between the Von Neumann entropy and
the changes in the density matrix due to a measurement.}

\begin{theorem}
The sum of the absolute squares of all the indistinction-amplitudes or
coherences $\alpha_{i}\alpha_{i^{\prime}}^{\ast}$ that are zeroed (i.e.,
decohered) in the L\"{u}ders mixture operation taking $\rho\left(
\psi\right)  $ to $\hat{\rho}\left(  \psi\right)  $ is the quantum logical
entropy $h\left(  \hat{\rho}\left(  \psi\right)  \right)  $.
\cite{ell:entropy}
\end{theorem}

The notions of classical and quantum logical entropy give the respective
measures of information based on the foundational idea of
information-as-distinctions.\footnote{The above classical cases, dealing with
sets instead of vectors, could be made even closer to QM by using a vector
space where each vector (represented in a basis set) is a set. That is the
case for vector spaces over $%
\mathbb{Z}
_{2}$. The above machinery from the classical cases formulated over $%
\mathbb{Z}
_{2}^{n}$ gives a pedagogical (or \textquotedblleft toy\textquotedblright)
model of QM--\textquotedblleft quantum mechanics over sets.\textquotedblright%
\ \cite{ell:qm-sets}}

\section{Quantum dynamics and measurement}

\subsection{Von Neumann's type 2 processes}

Von Neumann divided quantum processes into two fundamentally different types:

\begin{enumerate}
\item \textquotedblleft the arbitrary changes by
measurements,\textquotedblright\ and

\item \textquotedblleft the automatic changes which occur with passage of
time.\textquotedblright\ \cite[p. 351]{von-n:foundations}
\end{enumerate}

The OI interpretation needs to `make sense' out of these two different types
of processes in terms of distinctions and indistinctions. Indeed, the
difference is between:

\begin{enumerate}
\item processes that make distinctions, and

\item processes that preserve distinctions.
\end{enumerate}

Taking $\#2$ first, The degree to which two quantum states are indistinct or
distinct is given by the inner product $\left\langle \phi|\psi\right\rangle $,
so a quantum process that does not change this amplitude of indistinction
between states is mathematically described as a unitary transformation (i.e.,
a linear transformation that preserves inner products). The unitary evolution
of superpositions, e.g., $\left\vert \psi\left(  t\right)  \right\rangle
=U\left(  t,t_{0}\right)  \left\vert \psi\left(  t_{0}\right)  \right\rangle
$, is a mathematical description of the propagation of waves. The connection
between unitary transformations and the solutions to the Schr\"{o}dinger
\textquotedblleft wave\textquotedblright\ equation is given by Stone's Theorem
\cite{stone:thm}: there is a one-to-one correspondence between strongly
continuous $1$-parameter unitary groups $\left\{  U\left(  t,t_{0}\right)
\right\}  _{t\in%
\mathbb{R}
}$ and Hermitian operators $H$ on the Hilbert space so that $U(t,t_{0}%
)=e^{iHt}$.

In simplest terms, a unitary transformation describes a rotation such as the
rotation of a unit vector in the complex plane.%

\begin{center}
\includegraphics[
height=1.3681in,
width=3.9764in
]%
{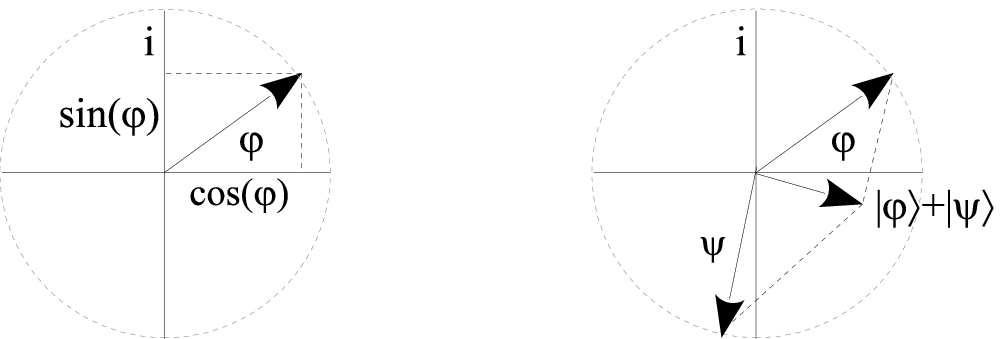}%
\end{center}

\begin{center}
Figure 2: Rotating vector and addition of vectors
\end{center}

\noindent The rotating unit vector traces out the cosine and sine
\textquotedblleft wave\textquotedblright\ functions on the two axes, and the
position of the arrow can be compactly described as a function of $\varphi$
using Euler's formula:

\begin{center}
$e^{i\varphi}=\cos\left(  \varphi\right)  +i\sin\left(  \varphi\right)  $.
\end{center}

\noindent Such complex exponentials and their superpositions are the
\textquotedblleft wave functions\textquotedblright\ of QM. The
\textquotedblleft wave functions\textquotedblright\ describe the evolution of
particles in indefinite states in isolated systems where there are no
distinction-creating interactions to change the degree of indistinctness
between states, i.e., the context where Schr\"{o}dinger's equation holds.
Classically it has been assumed that the mathematics of waves must describe
physical waves of some sort, and thus the puzzlement about the
\textquotedblleft wave functions\textquotedblright\ of QM having complex
amplitudes (in $3N$-dimensional space for systems of $N$ particles) and no
corresponding physical waves.

But we have supplied \textit{another} interpretation; wave mathematics is the
mathematics of indefiniteness-preserving evolution, i.e., superposition
represents indefiniteness and unitary evolution represents the
indistinctness-preserving evolution of an isolated system. The same
mathematics describes both types of evolution. Using the wave interpretation
instead of the indefiniteness interpretation of the mathematics has been one
of most historic wrong turns in the interpretation of QM--which has continued
long after it was realized that the \textquotedblleft wave
function\textquotedblright\ could not describe actual physical waves. But
humans have evolved so they can readily imagine the evolution of common
macro-phenomena such as the propagation of waves, while indefinite states and
their evolution present a much greater challenge to the imagination.

Richard Feynman's approach to QM shows how to develop the mathematics of QM
without appeal to waves (although wave imagery may be used as a pedagogical crutch).

\begin{quotation}
\noindent I want to emphasize that light comes in this form--particles. It is
very important to know that light behaves like particles, especially for those
of you who have gone to school, where you were probably told something about
light behaving like waves. I'm telling you the way it does behave--like
particles. \cite[p. 15]{feynman:qed}
\end{quotation}

\noindent Indeed, Feynman takes note of cases where the wave theory falls
short since:

\begin{quotation}
\noindent\noindent the wave theory cannot explain how the detector makes
equally loud clicks as the light gets dimmer. Quantum electrodynamics
\textquotedblleft resolves\textquotedblright\ this wave-particle duality by
saying that light is made of particles (as Newton originally thought), but the
price of this great advancement of science is a retreat by physics to the
position of being able to calculate only the probability that a photon will
hit a detector, without offering a good model of how it actually happens.
\cite[pp. 36-7]{feynman:qed}
\end{quotation}

\noindent The OI interpretation argues that what \textquotedblleft actually
happens\textquotedblright\ in \textquotedblleft wave-like\textquotedblright%
\ behavior is the evolution of a particle that is indefinite between a number
of undistinguished alternatives (a type $2$ process), and thus the OI
interpretation could be seen as attempting to give an ontology that underlies
Feynman's mathematical approach to QM. For instance, in the double slit
experiment, instead of saying \textquotedblleft the electron sweeps from
source to screen following all possible paths at once\textquotedblright%
\ \cite[p. 32]{cox-forshaw:q-universe}, it would be better to say that the
electron was in a state of being indefinite between all the possible paths in
going from source to screen. By developing the indefiniteness interpretation
to the superposition of paths in the Feynman approach, one has a realistic
non-wave interpretation of QM.

\subsection{Von Neumann's type 1 processes}

The $\#1$ type process is a process that does make distinctions. Richard
Feynman has given perhaps the clearest characterization of the two types of
processes in terms of distinctions and indistinctions.

\begin{quotation}
\noindent If you could, in principle, distinguish the alternative final states
(even though you do not bother to do so), the total, final probability is
obtained by calculating the probability for each state (not the amplitude) and
then adding them together. If you cannot distinguish the final states even in
principle, then the probability amplitudes must be summed before taking the
absolute square to find the actual probability. \cite[p. 3-16]%
{feynman:lectures3}
\end{quotation}

\noindent Feynman gives examples that do not involve any macroscopic measuring
apparatus (neutrons scattering in crystals or collisions of alpha-particles)
to avoid all the extraneous considerations (e.g., environmental dephasing) in
the literature on measurement. For instance, Feynman considers the case where
\textquotedblleft all neutrons from the source having spin up and all the
nuclei of the crystal having spin down\textquotedblright\ \cite[p.
3-15]{feynman:lectures3} If a scattered neutron has spin down, then one of the
atoms in the crystal must have spin up so the different paths through the
crystal are distinguished. That is a type $1$ process which makes distinctions
between the paths so the amplitude of each path is (separately) squared to
find its probability. If the alternatives cannot in principle be
distinguished, then it is a type $2$ process of unitary evolution of the
indefinite superposition of the paths, so the path amplitudes are added before
taking the absolute squares to determine the probability--which will then
reflect the interference between the paths.

One can extract from Feynman's probability rules the basic
\textit{distinguishability principle} that separates type $2$ unitary
evolution from the type $1$ state reduction or `measurement.' Consider the
unitary evolution of a particle in an indefinite state that is a superposition
of various definite states. If the particle then undergoes an interaction
where the outcomes of the superposed definite states can, in principle, be
distinguished, then the states \textit{are} distinguished and the particle
emerges from the interaction in one of the definite states with the
probability determined by the absolute square of its amplitude in the
superposition (Born rule). In short, if an interaction \textit{has to} make a
difference between the superposed states in the final outcomes, then it
\textit{does} make a difference in that the indefinite superposition state is
reduced to one of the definite states that were superposed.

Hermann Weyl likened a measurement to a particle having to pass through a
\textquotedblleft sieve or grating\textquotedblright\ \cite[p. 259]%
{weyl:phil}. For an intuitive image, think of a \textquotedblleft
blob\textquotedblright\ of dough as the indefinite superposition of a set of
polygonal shapes. The blob evolves as a blob until it hits a grating with
holes corresponding to the superposed shapes so the blob then has to pass
through one of the holes and thus gets `ontologically classified' as one of
the definite shapes. The grating distinguishes and classifies the shapes in
the indefinite superposition.%

\begin{center}
\includegraphics[
height=2.13in,
width=2.4491in
]%
{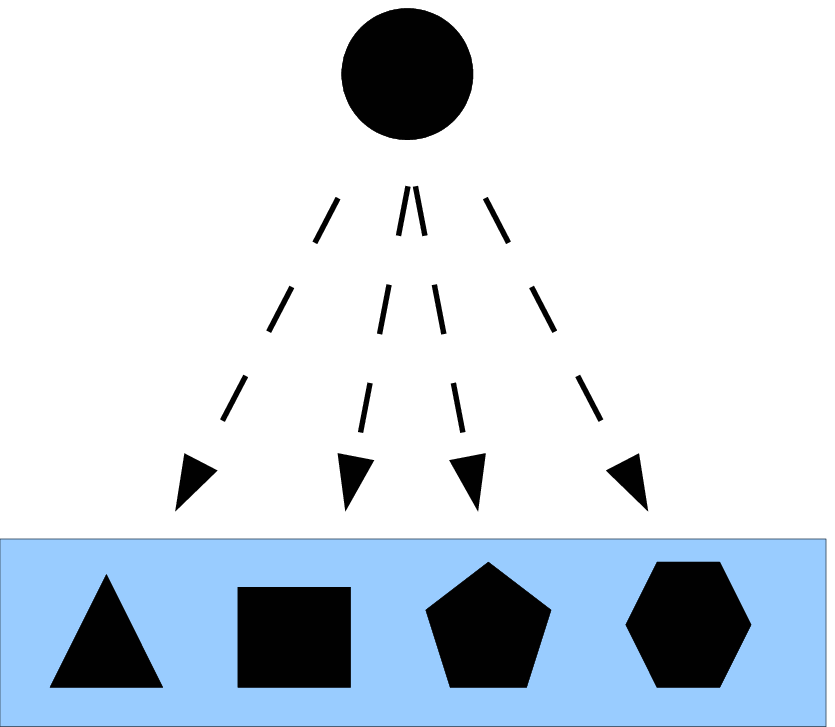}%
\end{center}

\begin{center}
Figure 3:\ Measurement where a superposition of definite shapes has to take on
one of the shapes.
\end{center}

Werner Heisenberg is usually presented as an advocate of the Copenhagen
interpretation of QM. But in his mature philosophical reflections, e.g.,
\cite{heisenberg:p-and-p}, he used the imagery of \textquotedblleft
potentiality\textquotedblright\ and \textquotedblleft
actuality\textquotedblright\ which, as noted by Shimony, can be interpreted as
\textquotedblleft indefiniteness\textquotedblright\ and \textquotedblleft
definiteness\textquotedblright\ respectively.

\begin{quotation}
\noindent Heisenberg \cite[p. 53]{heisenberg:p-and-p} used the term
\textquotedblleft potentiality\textquotedblright\ to characterize a property
which is objectively indefinite, whose value when actualized is a matter of
objective chance, and which is assigned a definite probability by an algorithm
presupposing a definite mathematical structure of states and properties.
Potentiality is a modality that is somehow intermediate between actuality and
mere logical possibility. That properties can have this modality, and that
states of physical systems are characterized partially by the potentialities
they determine and not just by the catalogue of properties to which they
assign definite values, are profound discoveries about the world, rather than
about human knowledge. It is fair to say, in view of my discussion above of
metaphysics, that these statements about quantum mechanical potentiality are
metaphysical propositions suggested by the formalism of quantum mechanics.
These statements, together with the theses about potentiality, may
collectively be called \textquotedblleft the Literal
Interpretation\textquotedblright\ of quantum mechanics. \cite[p.
6]{shim:vienna}
\end{quotation}

\noindent Heisenberg's use of the notion of \textquotedblleft
potentiality\textquotedblright\ in contrast to actuality does not seem
appropriate since the \textquotedblleft potentialities\textquotedblright\ have
very real effects on actuality (e.g., the quantum interference effects), so it
would seem more appropriate to consider indefinite and definite actualities.

\section{Group theory and QM}

We have argued that reality at the quantum level is inherently indefinite
which under certain circumstances becomes partly definite. The understanding
of quantum reality in terms of indistinguishability is already well-known in
the area of \textquotedblleft identical\textquotedblright\ particles, e.g.,
\cite{castellani:bodies}, and that will not be recapitulated here. We have
also emphasized the new light thrown on these questions by the new
developments in the logic of partitions (= equivalence relations = quotient
sets) and in its quantitative development as logical information theory. There
is another already well-known area of mathematics dealing the specification of
equivalences (or symmetries), namely, group theory--so one would expect it to
be highly applicable to QM. And it is.

An equivalence relation is a transitive, symmetric, and reflexive relation. A
group operating on a set is a natural way to define an equivalence on the set
(the partition of orbits) since a group operation is an associative operation
that is closed under composition (transitivity), has inverses (symmetry), and
includes the identity operation (reflexivity) [for more, see
\cite{brading:symmetries}].

To briefly touch on a quantum example, we need to lift or generalize the set
case of a group operating on a set, i.e., a set representation of the group
operations, to complex vector space representations of a (symmetry) group. As
noted above, a set partition generalizes to a direct-sum decomposition of a
vector space. The set partition of orbits generalizes to the direct-sum
decomposition of a complex vector space into irreducible subspaces. A
representation restricted to an irreducible subspace is an irreducible
representation. For a certain symmetry group of particle physics,
\textquotedblleft an elementary particle `is' an irreducible unitary
representation of the group.\textquotedblright\ \cite[p. 149]{stern:group}
Thus our approach from partitions and equivalence relations comports with
\textquotedblleft the soundness of programs that ground particle properties in
the irreducible representations of symmetry
transformations...\textquotedblright\ \cite[p. 171]{fine:shaky} (for more, see
\cite{ell:oid-interp}).

\section{Concluding remark}

One way to succinctly describe the objective indefiniteness interpretation of
QM is that the mathematics for the evolution of the quantum \textquotedblleft
wave function\textquotedblright\ is also the mathematics for the
indistinction-preserving evolution of indefinite (superposition) states. The
so-called \textquotedblleft wave-particle duality\textquotedblright\ is really
the juxtaposition of a particle evolving with an indefinite position
(\textquotedblleft wave-like\textquotedblright\ behavior) with a particle
having a definite position. The objective indefiniteness approach to
interpreting QM thus provides an explanation for the appearance of the
mathematics of waves (which implies interference as well as the quantized
solutions to the \textquotedblleft wave\textquotedblright\ equation that gave
QM its name) when, in fact, there are no actual physical waves involved.


\begin{thebibliography}{99}                                                                                               %


\bibitem {anil:twodoors}Ananthaswamy, Anil. 2018. \textit{Through Two Doors At
Once: The Elegant Experiment That Captures the Enigma of Our Quantum Reality}.
New York: Dutton.

\bibitem {auletta:qm}Auletta, Gennaro, Mauro Fortunato, and Giorgio Parisi.
2009. \textit{Quantum Mechanics}. Cambridge UK: Cambridge University Press.

\bibitem {baues:ht}Baues, Hans-Joachim. 1995. Homotopy Types. In
\textit{Handbook of Algebraic Topology}, edited by I. M. James, 1--72.
Amsterdam: Elsevier Science.

\bibitem {birkvon-n:logicqm}Birkhoff, Garrett, and John Von Neumann. 1936.
\textquotedblleft The Logic of Quantum Mechanics.\textquotedblright%
\ \textit{Annals of Mathematics} 37 (4): 823--43.

\bibitem {boole:lot}Boole, George. 1854. \textit{An Investigation of the Laws
of Thought on Which Are Founded the Mathematical Theories of Logic and
Probabilities}. Cambridge: Macmillan and Co.

\bibitem {brading:symmetries}Brading, Katherine, and Elena Castellani. 2003.
\textit{Symmetries in Physics: Philosophical Reflections}. Cambridge UK:
Cambridge University Press.

\bibitem {castellani:bodies}Castellani, Elena, ed. 1998. \textit{Interpreting
Bodies: Classical and Quantum Objects in Modern Physics}. Princeton: Princeton
University Press.

\bibitem {cohen-tann:qm}Cohen-Tannoudji, Claude, Bernard Diu, and Franck
Lalo\"{e}. 2005. \textit{Quantum Mechanics: Volumes 1 and 2}. New York: John
Wiley \& Sons.

\bibitem {cox-forshaw:q-universe}Cox, Brian, and Jeff Forshaw. 2011.
\textit{The Quantum Universe (and Why Anything That Can Happen, Does)}.
Boston: DaCapo Press.

\bibitem {ell:countingdits}Ellerman, David. 2009. \textquotedblleft Counting
Distinctions: On the Conceptual Foundations of Shannon's Information
Theory.\textquotedblright\ \textit{Synthese} 168 (1 May): 119--49.

\bibitem {ell:partitionlogic}Ellerman, David. 2010. \textquotedblleft The
Logic of Partitions: Introduction to the Dual of the Logic of
Subsets.\textquotedblright\ \textit{Review of Symbolic Logic} 3 (2 June): 287--350.

\bibitem {ell:intropartlogic}Ellerman, David. 2014. \textquotedblleft An
Introduction to Partition Logic.\textquotedblright\ \textit{Logic Journal of
the IGPL} 22 (1): 94--125.

\bibitem {ell:delayed}Ellerman, David. 2015. \textquotedblleft Why Delayed
Choice Experiments Do NOT Imply Retrocausality.\textquotedblright%
\ \textit{Quantum Studies: Mathematics and Foundations} 2 (2): 183--99.

\bibitem {ell:newfound}Ellerman, David. 2017. \textquotedblleft Logical
Information Theory: New Foundations for Information Theory.\textquotedblright%
\ \textit{Logic Journal of the IGPL} 25 (5 Oct.): 806--35. https://doi.org/10.1093/jigpal/jzx022.

\bibitem {ell:qm-sets}Ellerman, David. 2017. \textquotedblleft Quantum
Mechanics over Sets: A Pedagogical Model with Non-Commutative finite
Probability Theory as Its Quantum Probability Calculus.\textquotedblright%
\ \textit{Synthese} 194 (12): 4863--96.

\bibitem {ell:entropy}Ellerman, David. 2018. \textquotedblleft Logical
Entropy: Introduction to Classical and Quantum Logical Information
Theory.\textquotedblright\ \textit{Entropy} 20 (9): Article ID 679. https://doi.org/10.3390/e20090679.

\bibitem {ell:dsd-logic}Ellerman, David. 2018. \textquotedblleft The Quantum
Logic of Direct-Sum Decompositions: The Dual to the Quantum Logic of
Subspaces.\textquotedblright\ \textit{Logic Journal of the IGPL} 26 (1
January): 1--13. https://doi.org/10.1093/jigpal/jzx026.

\bibitem {ell:oid-interp}Ellerman, David. 2018. \textquotedblleft The
Objective Indefiniteness Interpretation of Quantum
Mechanics.\textquotedblright\ \textit{ArXiv.Org}. September 2018. http://arxiv.org/abs/1210.7659v2.

\bibitem {feynman:qed}Feynman, Richard P. 1985. \textit{QED: The Strange
Theory of Light and Matter}. Princeton NJ: Princeton University Press.

\bibitem {feynman:lectures3}Feynman, Richard P., Robert B. Leighton, and
Matthew Sands. 2010. \textit{The Feynman Lectures on Physics: Quantum
Mechanics Vol. III (New Millennium Ed.)}. Reading MA: Addison-Wesley.

\bibitem {finberg:rota}Finberg, David, Matteo Mainetti, and Gian-Carlo Rota.
1996. \textquotedblleft The Logic of Commuting Equivalence
Relations.\textquotedblright\ In \textit{Logic and Algebra}, edited by Aldo
Ursini and Paolo Agliano, 69--96. New York: Marcel Dekker.

\bibitem {fine:shaky}Fine, Arthur 1986. \textit{The Shaky Game: Einstein,
Realism, and the Quantum Theory}. Chicago: University of Chicago Press.

\bibitem {heisenberg:p-and-p}Heisenberg, Werner. 1962. \textit{Physics \&
Philosophy: The Revolution in Modern Science}. New York: Harper Torchbooks.

\bibitem {jaeger:2014}Jaeger, Gregg. 2014. \textit{Quantum Objects: Non-Local
Correlation, Causality and Objective Indefiniteness in the Quantum World}.
Heidelberg: Springer.

\bibitem {shannon:belljournal}Shannon, Claude E. 1948. \textquotedblleft A
Mathematical Theory of Communication.\textquotedblright\ \textit{Bell System
Technical Journal} 27: 379--423; 623--56.

\bibitem {shapiro:cut}Shapiro, Stewart. 2011. The Company Kept by
Cut-Abstraction (and It Relatives). \textit{Philosophia Mathematica (III)} 19: 107--38.

\bibitem {shim:reality}Shimony, Abner 1988. The reality of the quantum world.
\textit{Scientific American}. 258 (1): 46-53.

\bibitem {shim:vienna}Shimony, Abner. 1999. \textquotedblleft Philosophical
and Experimental Perspectives on Quantum Physics.\textquotedblright\ In
\textit{Philosophical and Experimental Perspectives on Quantum Physics: Vienna
Circle Institute Yearbook 7}, 1--18. Dordrecht: Springer Science+Business Media.

\bibitem {stern:group}Sternberg, Shlomo 1994. \textit{Group Theory and
Physics}. Cambridge: Cambridge University Press.

\bibitem {stone:thm}Stone, Marshall H. 1932. On one-parameter unitary groups
in Hilbert Space. \textit{Annals of Mathematics}. 33 (3): 643--648.

\bibitem {ufp:hott}Univalent Foundations Program. 2013. \textit{Homotopy Type
Theory: Univalent Foundations of Mathematics}. Princeton: Institute for
Advanced Studies.

\bibitem {von-n:foundations}Von Neumann, John. 1955. \textit{Mathematical
Foundations of Quantum Mechanics}. Translated by Robert Beyer. Princeton:
Princeton University Press.

\bibitem {weyl:phil}Weyl, Hermann. 1949. \textit{Philosophy of Mathematics and
Natural Science}. Princeton: Princeton University Press.
\end{thebibliography}
\end{document}